\documentclass[conference]{IEEEtran}

\usepackage[cmex10]{amsmath}
\usepackage{cite}
\usepackage{amssymb}
\usepackage{color}
\usepackage{graphicx}
\newtheorem{teorema}{Theorem}
\newtheorem{proposicao}{Proposition}
\newtheorem{remark}{Remark}
\newtheorem{lema}{Lemma}
\newtheorem{definicao}{Definition}

\newtheorem{exemplo}{Example}

\newcommand{\F}{\mathbb{F}}

\newcommand{\Fqn}{\mathbb{F}_q^n}

 \title{Weights which respect support and NN-decoding}
\author{
 \IEEEauthorblockN{Roberto Machado and Marcelo Firer}
 \IEEEauthorblockA{Institute of Mathematics, Statistics and Scientific Computing\\
 	University of Campinas\\
 	Email: robertomachado@ime.unicamp.br, mfirer@ime.unicamp.br}}

\begin{document}
\maketitle
\begin{abstract}
	In this work we explore a family of metrics over finite fields which respect the support of vectors. We show how these metrics can be obtained from the edge-weighted Hamming cube and, based on this representation we give a description of a  group of linear isometries (with respect to the metric). Next we introduce the concept of conditional sum of metrics and determine what conditions determine a metric respecting support, out of two such given metrics. Finally we introduce the labeled-poset block metrics, a new family of metrics which respects support of vector, filling a gap existing in the known such metrics. For this family we give a full description of the group of linear isometries and determine necessary and sufficient conditions for the existence of a MacWilliams identity.
\end{abstract}

\section{Introduction}\label{sec:intro}

In coding theory, there are two main sources of decoding criteria: a probabilistic (Maximum Likelihood Decoding - MLD) and a metric (Minimum Distance Decoding - MDD). While the first one focuses on the properties of the channel and is the optimal criterion (in term of minimizing the error probability of the encoding-transmission-decoding process), the last generally has properties that may help in the implementation of decoding algorithms.

The most important instance of channel is the binary symmetric channel
which MLD criterion matches the MDD criterion determined by the Hamming metric.
The Hamming metric $d_H$ has two important properties that are very valuable:

\textbf{P1} \textit{\underline{ Weight condition:}} The metric $d_H$ is determined by the Hamming weight $\mathrm{wt}_H$, i.e., $d_H(u,v)=\mathrm{wt}_H(u-v)$.

\textbf{P2}  \textit{\underline{  Support condition:}} If the vectors $u=(u_1,\ldots,u_n)$ and $u^{\prime}=(u^{\prime}_1,\ldots,u^{\prime}_n)$
are such that $u_i\neq 0$ whenever $u^{\prime}_i\neq 0$, then $\mathrm{wt}_H(u)\geq \mathrm{wt}_H(u^{\prime})$. In this case we say that the metrics \emph{respects support}.

The first of this properties gives an important tool for implementing algorithms: the well known syndrome decoding may be performed for every metric determined by a weight.  The second property  makes it meaningful in the context of coding theory, in the sense that making extra errors can not improve the result. Altogether, we name these as our \textit{basic decoding conditions} (BDC, for short).

In the literature, matching between channels and metrics (that is,  the maximum likelihood decoding coincides with nearest neighbor decoding) is not much explored. Despite the large number of channels that are studied and the large number of metrics described in the literature in the context of Coding Theory (see, for example, [Chapter 16] in \cite{deza},  and a recent survey of Gabidulin \cite{gabidulin2012brief}), there are a few examples of classical metrics and channels which are proved to be matched.

Although matching channels and metrics is not widely studied, we can find in the literature large families of metrics satisfying the basic decoding conditions. We cite, for example, the poset metrics of Brualdi \cite{brualdi}, Gabidulin's combinatorial metrics\cite{gabidulin}, poset-block metrics \cite{muniz} and digraph metrics \cite{etzion}. \footnote{Let $[n]=\{1,2,...,n\}$. Given a partial order (poset) $P=([n],\preceq_P)$ on the set $[n]$, the poset weight $\mathrm{wt}_P(x)$ of a vector $x\in\mathbb{F}_q^n$ is defined as $| \langle \mathrm{supp}(x)\rangle_P |$, where $\mathrm{supp}(c)$ is the support of the vector, $\langle X \rangle_P$ is the smallest order-ideal containing $X$ and $|A|$ is the cardinality of $A$. If $\mathcal{F}=\{A_1,A_2,\cdots A_r\}$ is a covering of $[n]$, the $\mathcal{F}$-combinatorial weight of a vector $x\in\mathbb{F}_q^n$ is $\min \{|\mathcal{A}|: \mathcal{A}\subset \mathcal{F} \text{ and } \mathcal{A} \text{ is a covering of } \mathrm{supp}(x)  \}$. For more details on all those metrics, see \cite{livro}.
}   

All  those generalize the Hamming metric and they represent very large families of metrics over a vector space $\mathbb{F}_q^n$ (large in the sense that each of those families grows exponentially with $n$). Nevertheless, those are not sufficient to determine all the MDD criteria satisfying the support condition. Example \ref{example1} illustrates such an affirmation for the smallest possible case, $n=2$.

Before introducing the example, we should remark that different metrics may determine the same MMD, and in this case, we should consider such metrics to be equivalent. To be more precise: two metrics $d_1$ and $d_2$ over a space $V$ are \textit{decoding-equivalent} if given any code $C\subset V$ and any received message $x\in V$, the MDDs determined by both metrics generate the same set of codewords, i.e., $\arg\min_{c\in C}d_1(x,c) =\arg\min_{c\in C}d_2(x,c) $, for all $x\in\mathbb{F}_q^n$. It is not difficult to prove that for metrics defined by weights, being equivalent means that, when ordering the  vectors in $\mathbb{F}_q^n$ according to the two different weights we get the same ordering  (see \cite{rafa} for details). 



\begin{exemplo}\label{example1}
	Let us consider the space $\mathbb{F}_2^2=\{00, 10, 01, 11\}$. In this case we have 4 non decoding-equivalent criteria. In the table bellow we present these criteria and check each that can be determined by a metric in one of the large families we have mentioned: poset $\mathrm{wt}_P$, poset-blocks $\mathrm{wt}_{PB}$, combinatorial $\mathrm{wt}_C$ and digraph $\mathrm{wt}_D$. It is worth to note that only the first one can be determined by any of this families of metrics.
	
	
	\begin{table}[h]
		\begin{flushleft}
			\begin{tabular}{|c|c|c|c|c|c|}
				\hline 
				Criterion	& {\footnotesize$\mathrm{wt}_H$} & {\footnotesize$\mathrm{wt}_{P}$ }& {\footnotesize$\mathrm{wt}_{PB}$} & {\footnotesize$\mathrm{wt}_{C}$} &  {\footnotesize$\mathrm{wt}_D$}\\ 
				\hline 
				{\footnotesize 	$\mathrm{wt}(10)=\mathrm{wt}(01)<\mathrm{wt}(11)$}	&  \checkmark& \checkmark & \checkmark &  \checkmark& \checkmark\\ 
				\hline 
				{\footnotesize	$\mathrm{wt}(10)=\mathrm{wt}(01)=\mathrm{wt}(11)$}	&  &  &  \checkmark&  \checkmark&\checkmark\\
				\hline 
				{\footnotesize	$\mathrm{wt}(10)<\mathrm{wt}(01)=\mathrm{wt}(11)$}	&  & \checkmark &  \checkmark&  &\checkmark\\
				\hline 
				{\footnotesize	$\mathrm{wt}(10)<\mathrm{wt}(01)<\mathrm{wt}(11)$}	&  &  &  &  &\\
				\hline 
			\end{tabular}\\
			\vspace{3pt}
			\caption{Decoding criteria which respect support in $\mathbb{F}_2^2$}\label{table}
		\end{flushleft} 
	\end{table}
	
	\vspace{-20pt}
	%
	%
	%
	%
	
	We stress that different metrics can be decoding-equivalent.  In fact, even though the second criterion in the table ($\mathrm{wt}(u)  $ is constant for $u\neq 0$) may be determined  by  a poset-block  $\mathrm{wt}_{PB}$ and also by a digraph $\mathrm{wt}_{D}$ weight, 
	simple computations shows that $\mathrm{wt}_{PB}(u)=1$ and $\mathrm{wt}_{D}(u)=2$, for $u\neq 0$. More important, we note that the last decoding criterion can not be determined by any metric belonging to one of these families.
\end{exemplo}

\medskip

This work aims to reduce the gap between the known and studied metrics satisfying the BDC and the space of all possible metrics satisfying the BDC. 

\medskip
In Section \ref{metrics} we give the first systematic approach to the space of all metrics satisfying the BDC. We introduce a conditional operator on metrics and the main result is to establish what are the conditions that ensure that from a pair of metrics satisfying the BDC one gets another metric satisfying the BDC. This is a starting point to estimate how large the metrics obtained by a conditional sum of a poset, digraph and combinatorial metrics is in the space of a metrics satisfying the BDC. In other words, this section moves one step forward in a long term goal to develop  an ``approximation theory" of metrics in the context of coding theory.

The rest of the work is devoted to introduce a new family of metrics that generalizes both the digraph metrics and the poset-block metrics, introduced in  \cite{etzion} and \cite{muniz}, respectively.  In Section \ref{sec:preliminaries}, we introduce the basic concepts and define the labeled-poset-block metrics. In Section  \ref{isometriess}  we characterize  the group of linear isometries of a space endowed with a labeled-poset-block metric. In Section \ref{candec} we give necessary and sufficient conditions to ensure that every linear code admits a canonical decomposition and derive necessary condition to ensure the existence of a MacWilliams' Identity.

\section{Operating with metrics which respect support}\label{metrics}

Since we are concerned with metrics determined by weights, we establish a condition on weights to ensure that the corresponding metric satisfies the MDC.

\begin{definicao}
	A function $\mathrm{wt} :\mathbb{F}_2^n \rightarrow \mathbb{Z}$ is a \textit{weight respecting support} (or simply S-weight) if the following holds:
	
	\begin{enumerate}
		\item $\mathrm{wt}(u)\geq 0$ and equality implies $x=0$;
		\item If $\mathrm{supp}(u) \subset \mathrm{supp}(v)$, then $\mathrm{wt}(u)\leq \mathrm{wt}(v)$, where $\mathrm{supp}(x)=\{i\in \{1,\ldots, n\}:x_i\neq0\}.$
	\end{enumerate}
\end{definicao}

An S-weight  determines a semi-metric, by defining $d(u,v)=\mathrm{wt}(u-v)$, and two weights determine the same semi-metric if, and only if, they are equal. In order to guarantee the $d(u,v)=\mathrm{wt}(u-v)$ is a metric it is required the triangle inequality which we ignore this work due to the fact that every semi-metric $d$ can be rescaled to a decoding-equivalent metric $d'$ as follows:
\[ d'(u,v) = 
\left\{\begin{array}{cc}
	0,&\mbox{if}\quad u=v,\\
	 d(u,v)+ \max_{x,y\in \mathbb{F}_q^n} d(x,y), &\mbox{if}\quad u\neq v.
\end{array}\right.
\]
So, to understand the space of all metrics satisfying BDC (conditions P1 and P2), it is enough to study the space of all S-weights up to the following equivalence:

\begin{definicao}
	We say that two S-weights $\mathrm{wt}_1$ and $\mathrm{wt}_2$ are \emph{equivalent} (denoted by $\mathrm{wt}_1\sim \mathrm{wt}_2$ ) if  \[\mathrm{wt}_1(u)<\mathrm{wt}_1(v) \iff \mathrm{wt}_2(u)<\mathrm{wt}_2(v), \forall u,v\in \mathbb{F}_q^n .\]
\end{definicao}

It is not difficult to see that two S-weights are equivalent if, and only if, they are  decoding-equivalent  (in the sense defined in Section \ref{sec:intro}, see \cite{rafa} for details).

Given these initial definitions, we split this section into two parts. Through the first part, the $S$-weight functions are naturally represented by an edge-weighted Hamming cube that allows us to give a partial description for the group of linear isometries. We introduce the conditional sum of known weights in order to generate new $S$-weight which are not decoding-equivalent to the former ones and are able, for example, to fill the gap in the last row of Table \ref{table}.

We give necessary and sufficient conditions to guarantee that a conditional sum of two $S$-weights is also an $S$-weight. In addition, we prove that every $S$-weight is equivalent to an $S$-weight obtained by a  finite conditional sum of poset, digraph and combinatorial weights (to be introduced in ). 

\subsection{Weights respecting support}\label{graphs}

In order to explore properties of $S$-weights, our approach is to construct general $S$-weights in a way it can inherit the knowledge accumulated about poset, digraph and combinatorial metrics. For this purpose, we shall represent $S$-weights by labeling the edges of the Hamming cube.  This approach allows us to obtain a partial description for the group of linear isometries which is a fundamental tool in the context of coding theory. Indeed, being in the same orbit of this group is the definition of code equivalence and the structure of these groups is used to determine whenever the MacWilliams' extension property is satisfied and so forth.

We start by considering the Hamming cube $\mathcal{H}^n$ as a directed graph  where $\mathbb{F}_2^n$ is the set of vertices and $(u,v)$ is a (directed) arc  if, and only if, $d_H(u,v)=1$ and $\mathrm{wt}_H(u)<\mathrm{wt}_H(v)$. In addition, to every arc is assigned a non-negative integer $\delta(u,v)$. The pair $(\mathcal{H}^n, \delta) $ is called a \emph{$\delta$-weighted Hamming cube}. 
 
It is simple to see that given an $S$-weight $\mathrm{wt} :\mathbb{F}_q^n \rightarrow \mathbb{N}$, by setting $\delta((u,v)) = \mathrm{wt}(v)-\mathrm{wt}(u)$ we have that $\mathrm{wt}(w) = \sum_{(u,v)\in \tau}\delta((u,v))$ where $\tau$ is any \textit{trail} from the null vector to $w$.  By simplicity, we denote $\delta(\tau)=\sum_{(u,v)\in \tau}\delta((u,v))$ whenever no confusion may arise. This shows that every $S$-weight can be represented by a $\delta$ (weighted) Hamming cube. From here on, except if explicitly stated, we assume that every trail has $0$ as its initial vertex.



However, not every $(\mathcal{H}^n,\delta  )  $ determines an $S$-weight. For this to happen, the $\delta$ function should avoid the situation depicted in the following example. Consider the $\delta$-Hamming cube below.
\begin{center}
	\includegraphics[width=2.5cm]{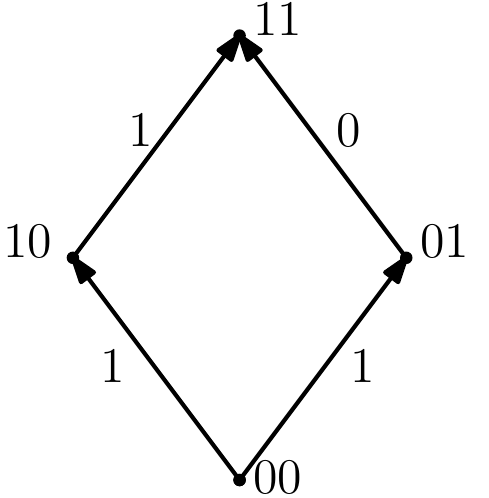}
\end{center}
This $\delta$ function does not induce an $S$-weight because the sum of weights on the left and right trails from $00$ to $11$ are different,  i.e., $\delta((00,10)) + \delta((10,11)) \neq \delta((00,01)) + \delta((01,11)) $. We avoid this situation imposing it as a necessary condition:

\begin{lema}\label{metric}
	The map  $\mathrm{wt}(w) = \delta(\tau)$, where $\tau$ is a trail from $0$ to $w$, is an $S$-weight if, and only if,   $\delta(0,e_i)>0$, for every $i\in\{1,\ldots,n\}$  and $\delta(\tau) = \delta(\tau^\prime)$, for any trails $\tau,\tau^\prime$ in $\mathcal{H}^n$ with same initial and final vertices. 
\end{lema}
\begin{IEEEproof} The proof follows directly from the definitions and it is omitted due to lack of space.
\end{IEEEproof}
The next proposition characterizes the weight functions which determines a combinatorial metric, as introduced by Gabidulin, \cite{gabidulin}.
\begin{proposicao}
	Let $\delta$ be determined by an $S$-weight $\mathrm{wt}$ as before. Then, $d(u,v)=\mathrm{wt}(u-v)$ is a combinatorial metric if, and only if,   $\delta((u,v)) \in \{0,1\}$.
\end{proposicao}
\begin{IEEEproof}
	If $\delta((u,v)) \geq 2$, then $\mathrm{wt}(v) = \mathrm{wt}(u) + \delta((u,v)) \leq \mathrm{wt}(u) + \mathrm{wt}(v-u)$. This implies that $\mathrm{wt}(v-u)\geq \delta((u,v)) \geq 2$. Since $v-u = e_i$, for some $i\in [n]$, and $\mathrm{wt}(e_i)=1$ for any combinatorial weight, we have that $\mathrm{wt}$ does not induce a combinatorial metric.	
	The proof of the \emph{if} part is constructive and will be omitted due to lack of space.
	%
\end{IEEEproof}



We reserve the rest of this work to present a standard form for a weight $\mathrm{wt}$, which will be used to prove some key coding results such as characterizing the group of linear isometries and the existence of a MacWilliams' Identity.
\begin{definicao}
	Let $\delta$ be obtained from an $S$-weight $\mathrm{wt}$. We say that $\delta$ is in a \textit{standard form} if, given a trail $\tau$ with $\delta(\tau)=k>1$ there is a trail $\tau^\prime$ such that $\delta(\tau^\prime)=k-1$. We say that an $S$-weight $\mathrm{wt}$ \emph{admits a standard form} if it is equivalent to an $S$-weight which determines a weight in a standard form.
\end{definicao}

\begin{proposicao}
	Every $S$-weight has a unique standard form.
\end{proposicao}
\begin{IEEEproof}
	The proof is obtained by following an algorithm which assign new values for $\delta$ which is a simple extension of the steps we detail in Example \ref{algorithm}.
\end{IEEEproof}
\begin{exemplo}\label{algorithm}
Consider the figure bellow. The $\delta$-Hamming cube on the left is not in a standard form since  $\delta(00, 01)=3$ and there is no trail $\tau$ with $\delta(\tau)=2$. In the middle, we assign the value $\delta((00, 01))=2$, and, since $\delta((00,10))+\delta((10,11)) = 4$, Lemma \ref{metric} imposes $\delta((01, 11))=2$ to get a weight wich defines an $S$-weight.  Now, on the right side, we repeat the procedure for the trail $\tau = \{(00,01),(01,11)\}$ which has $\delta(\tau)=4$ while there is no trail $\tau^\prime$ with $\delta(\tau)=3$. 
\begin{center}
		\includegraphics[width=8.9cm]{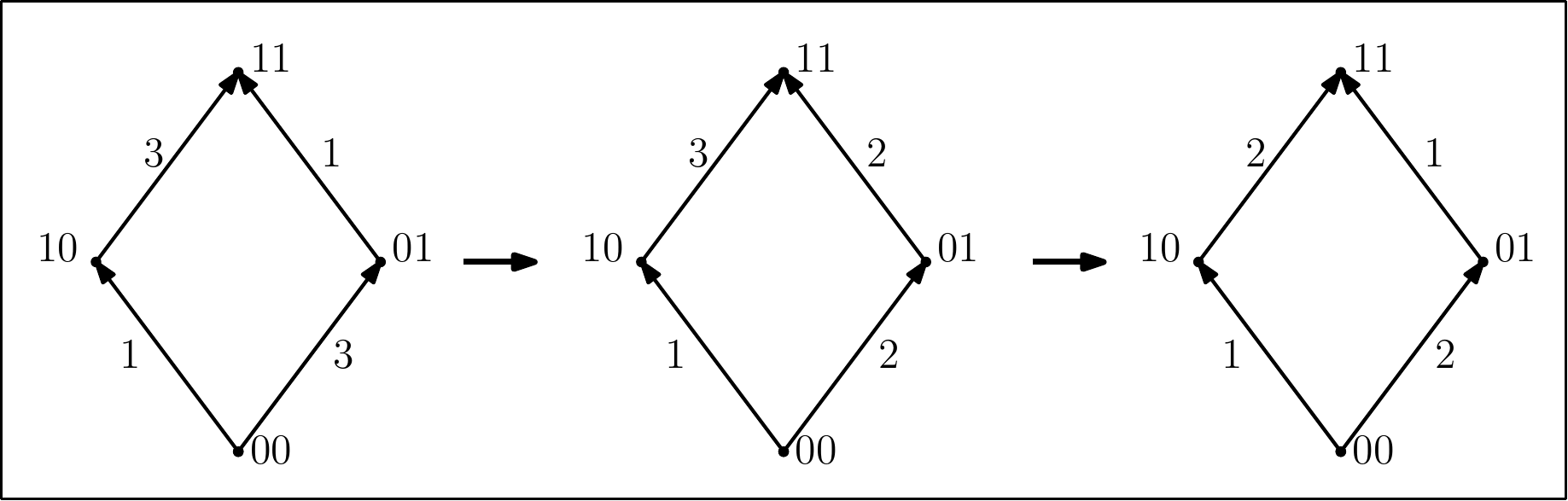}
\end{center}
We remark that since the values of $\delta$ decrease at every step, the algorithm will have a stopping point.
\end{exemplo}

From here on, we assume that every $\delta$ is in a standard form.  

Now, we turn our attention to describe the group of linear isometries in a space $\mathbb{F}_q^n$ endowed with a metric determined by an $S$-weight $\mathrm{wt}$, i.e.,
\begin{align*}
GL(n, q, \mathrm{wt}) = \{T:\mathbb{F}_q^n\rightarrow \mathbb{F}_q^n: T&\text{ is linear},\\& d(u,v)=d(T(u),T(v))\}.
\end{align*}
 Given  a linear map $T\!:\! \mathbb{F}_q^n\! \rightarrow \!\mathbb{F}_q^n$, we say it  \emph{respects domination} if: \emph{(i)}  $T(e_i) = \sum_{j=1}^{n}\lambda_j e_j$, with $\lambda_i\neq 0$; \emph{(ii)}  $\delta(\tau)=0$ if $\tau$ is a trail from $u\in \F_q^n$ to $u + T(e_i) - e_i$,  with $u_i\neq0$. We denote by $N(\mathcal{H},\delta)$ the group of transformations preserving domination and by $Aut(\mathcal{H},\delta)$ the group of automorphisms of $\mathcal{H}$ which preserve $\delta$.

The group $GL(n, q, \mathrm{wt})$ is fully described by the following propositions, which proofs are omitted due lack of space.

\begin{proposicao}
	Let $\phi\in S_{n}$.  The map $T_{\phi}:\mathbb{F}_q^n\rightarrow \mathbb{F}_q^n$, defined by $T_{\phi}(x_1, \ldots, x_n) = (x_{\phi(1)}, \ldots, x_{\phi(n)})$ is a linear isometry if, and only if, $T_\phi\in Aut(\mathcal{H},\delta)$. 
\end{proposicao}


\begin{proposicao}
	Any map $T\in N(\mathcal{H},\delta)$ is a linear isometry. 
\end{proposicao} 

\begin{teorema}
		Let $Aut\left(\delta\right)  \ltimes N\left( \delta\right)$ be the semi-direct product.  $Aut\left(\delta\right)  \ltimes N\left( \delta\right)\subseteq GL(n, q, \mathrm{wt})$ and the equality holds for every $q>2$.
\end{teorema}
%
%
%
%

\subsection{Conditional sums} \label{suba}

In previous section we showed how $S$-weights are related to $\delta$-Hamming cubes in the most general setting. The $S$-weights are not studied in its most general setting, so we wish to approximate a general or obtain a general $S$-weights from particular families that are better understood, for the same reason that one approximate smooth functions by polynomials. The most studied such metrics are the families of poset and combinatorial metrics, It is worth to note that they are virtually complementary, in the sense that the Hamming metric is the unique metric that belongs to both the families. 

So, we are left with two fundamental questions to be addressed in this section: How can we obtain a new $S$-weight out from two given $S$-weights? How large is the family of $S$-weights that can be constructed from a combination of poset and combinatorial weights? 

We start by presenting a conditional sum which permits to obtain new $S$-weights out of given ones. 

\begin{proposicao}
	Let $\mathrm{wt}_1$ and $\mathrm{wt}_2$ be $S$-weights. Then, the \textit{$k$-sum}
 \[\!(\mathrm{wt}_1 \oplus_k\mathrm{wt}_2)(u)\!=\!\left\{\begin{array}{rc}
\mathrm{wt}_1(u),&\!\mbox{if}\!\quad\!\mathrm{wt}_1(u)\!<\!k,\\
\!\mathrm{wt}_1(u)\!+\!\mathrm{wt}_2(u),&\mbox{if}\quad\!\mathrm{wt}_1(u)\!\geq\!k.
\end{array}\right.\] is a $S$-weight. 
\end{proposicao}
\begin{IEEEproof}
	The proof follows straightforward from definitions.	
\end{IEEEproof}
We remark that for $k=0,1$,  $\mathrm{wt}_1 \oplus_k\mathrm{wt}_2$ is the usual sum $\mathrm{wt}_1 + \mathrm{wt}_2   $. 
The previous proposition implies that the set of all $S$-weights endowed with directed sum or $k$-sum is a magma, i.e., the set of all weights is closed under ``$\oplus_k$".
\begin{remark}
	We stress that the conditional sum $\oplus_k$ may be replaced by a similar sum in which we consider different conditions \emph{respecting the support}\footnote{A condition $C$ on vectors of $\mathbb{F}_q^n$ \textit{respects support} if $C(u)=\texttt{true}$ implies $C(v)=\texttt{true}$ for every $v$ such that  $\mathrm{supp}(u) \subset  \mathrm{supp}(v)$. } of vectors $u\in\Fqn$. For instance, given two weights $\mathrm{wt}_1$ and $\mathrm{wt}_2$, let us consider the $(H,k)$-conditional sum $\mathrm{wt}_1\oplus_{(H,k)}\mathrm{wt}_2$ defined as follows
	
	$ \hspace{-6pt}(\mathrm{wt}_1 \!\oplus_{(H,k)} \!\mathrm{wt}_2)(u)\!=\! \hspace{-1pt} \left\{ \hspace{-3pt}\begin{array}{rc}
	\mathrm{wt}_1(u),&\!\text{if}\!\!\quad \mathrm{wt}_H(u)\!<\!k,\\
	\mathrm{wt}_1(u)+\mathrm{wt}_2(u), &\! \!\text{if}\! \quad \mathrm{wt}_H(u)\!\geq\!k.
	\end{array}\right.$
\end{remark}
%

\begin{exemplo}
Let $\mathrm{wt}_P$ be the poset weight that covers the criterion $\mathrm{wt}(00)<\mathrm{wt}(10)<\mathrm{wt}(01)=\mathrm{wt}(11)$ and $\mathrm{wt}_H$ be the Hamming weight. Then, the $(H,1)$-conditional sum $\mathrm{wt}_P\oplus_{(H,1)}\mathrm{wt}_H$ covers the remaining criterion in $\mathbb{F}_2^2$ given by $\mathrm{wt}(00)<\mathrm{wt}(10)<\mathrm{wt}(01)<\mathrm{wt}(11)$, the last row of Table \ref{table}.
\end{exemplo}
Not every conditional sum leads to non-equivalent metric. We wish to know under what conditions we have that  $\mathrm{wt}_1$, $\mathrm{wt}_2$ and  $\mathrm{wt}_1\oplus_C\mathrm{wt}_2$ are all equivalent. It happens, for example, for $\mathrm{wt}\oplus_0\mathrm{wt}$. We start with the following lemma:


\begin{lema}
	Let $\mathrm{wt}_1$ and $\mathrm{wt}_2$ be equivalent weights. Suppose that $\mathrm{wt}_1 \oplus_C \mathrm{wt}_2$ is also equivalent to $\mathrm{wt}_1$ and $\mathrm{wt}_2$ for a given condition $C$. Then, 
	\begin{enumerate}
		\item If $\mathrm{wt}_1(u) = \mathrm{wt}_2(v)$ and $C(u)=\texttt{true}$ then $C(v)=\texttt{true}$.  
		\item\label{cond} If $\mathrm{wt}_1(u) < \mathrm{wt}_2(v)$ and $C(u)=\texttt{true}$, then either $C(v)=\texttt{true}$  or $2\cdot \mathrm{wt}_1(u) < \mathrm{wt}_2(v)$.
	\end{enumerate}	
\end{lema}
\begin{IEEEproof}
	The proof can be obtained by simple computations and it is omitted.	
\end{IEEEproof}

The second part of the previous Lemma ensures that, if $\mathrm{wt}_1\sim \mathrm{wt}_2\sim \mathrm{wt}_1\oplus_C \mathrm{wt}_2  $ we can choose all of them to be equal and we get the following:

\begin{proposicao}
	Let $\mathrm{wt}$ be a $S$-weight. Then, $\mathrm{wt}\sim \mathrm{wt} \oplus_C \mathrm{wt}$  if, and only if, $\oplus_C =\oplus_k$, for some $k\in\mathbb{N}$.
\end{proposicao}
\begin{IEEEproof}
	It is enough to choose $k = \min\{\mathrm{wt}(u): u\in \mathbb{F}_q^n \text{ satisfies } C\}$.
\end{IEEEproof}

\begin{teorema}
	Every $S$-weight can be reached by a finite conditional sum of poset and combinatorial weights.
\end{teorema}
\begin{IEEEproof}
	The proof is omitted due to lack of space.
\end{IEEEproof}	
\section{Labeled-poset-block metrics}\label{sec:preliminaries}

In this section, we introduce a particular family which is a generalization for the digraph metrics which arises naturally from the reduced canonical form for directed graphs presented in \cite{etzion}. This canonical form makes a contraction of each maximal cycle into a unique vertex and, then, such vertex is labeled by the number of vertices contained in the original cycle. If we allow this labeling to assume different values such extension also generalizes the poset-block metrics by labeling every maximal cycle with 1. The goal in this section, is to present an structured family of metrics that covers the remaining decoding criterion in Table \ref{table}. Despite the generality of the approach, we also produce a description of group of linear isometries and determine conditions for a MacWilliams identity to be available.

Let $P=([m],\preceq_P)$ be a partially ordered set (abbreviated as poset), where $\preceq_p$ is a partial order over $[m]:=\{1,\ldots,m\}$. An \textit{ideal} in $P=([m],\preceq_P)$ is a subset  $I\subseteq [m]$ such that, if  $b\in I$ and $a \preceq_P b$, then $a\in I$.  Given 
$A\subseteq [m]$, we denote by $\langle A\rangle_P$ the smallest ideal of $P$ containing $A$ and call it as the \emph{ideal generated by} $A$.
An element $a$ of a set $A\subseteq [m]$ is called a \emph{maximal} element of $A$ if $a\preceq_P b$ for some $b\in A$ implies $b=a$. The set of all maximal elements of $A$ is denoted by $\mathcal{M}_P(A)$. Note that if $I\subseteq [m]$ is an ideal, then $\mathcal{M}_P(I)$ is the minimal set that generates $I$, i.e., $\langle \mathcal{M}_P(I)\rangle_P=I$. 

Given two posets $P$ and $Q$ over $[m]$, a \emph{poset isomorphism} is a bijection $\phi :[m]\rightarrow [m]$ such that $i\preceq_P j \! \iff \! \phi(i)\preceq_Q\phi(j)$. When $P=Q$, $\phi$ is called a $P$\emph{-automorphism}. The set of all $P$-automorphisms is a group denoted by $Aut(P)$.

A \textit{chain} in a poset $P$ is a subset $X\subseteq[m]$ such that 
any two elements $a,b\in X$ are comparable, in the sense that $a\preceq_Pb$ or $b\preceq_Pa$. We remark that any (finite) chain has a unique maximal element. The \emph{height}  $h(a)$ of an element $a \in P$ is the cardinality of a largest chain having  $a$ as the maximal element. The \emph{height} $h(P)$ \emph{of the poset} is the maximal height of its elements, i.e., $h(P)=\max\left\{h(a):a\in [m]\right\}  $. The $i$\emph{-th level} $\Gamma_i^P$ of a poset $P$ is the set of all elements with height $i$, i.e., $\Gamma_i^P = \{a\in [m] : h(a)=i\}.$
A poset $P$ is \emph{hierarchical} if elements at different levels are always comparable, i.e.,   $a\in \Gamma_i^P$ and $b\in \Gamma_{j}^P$ implies $a\prec_P b$ for any $1\leq i <j \leq h(P)$.

Let us consider a map $\pi : [n] \rightarrow [m]$ with $n\geq m$ (called a \textit{block map}). A vector $u \in\mathbb{F}_q^n$ may be written as $u= (u_1, \ldots, u_m)$, where $u_i\in \mathbb{F}_q^{k_i}$, with $k_i=|\pi^{-1}(i)|$. The $\pi$-\textit{support} is defined as 
$$\mathrm{supp}_{\pi}(u)=\{i\in [m]:u_i\neq0\}.$$

Given a block function $\pi: [n] \rightarrow [m]$,  a poset $P=([m],\preceq_P)$ and a \textit{label function} $L: [m] \rightarrow \mathbb{N}$, the $(P, \pi, L)$-\textit{weight} of $u$ is  defined as $$\mathrm{wt}_{(P, \pi, L)}(u)=\sum_{i\in\langle\mathrm{supp}_{\pi}(u)\rangle_P} L(i).$$
For $u,v\in \mathbb{F}_q^n$, we define the  \textit{labeled-poset-block distance} by:
$$d_{(P,\pi,L)}(u,v) = \mbox{\upshape{wt}}_{(P, \pi, L)}(u-v).$$

\begin{proposicao}
	If the label function $L$ assumes only positive values, then $d_{(P,\pi,L)}(u,v)$ determines a metric over $\Fqn$.
\end{proposicao}
\begin{IEEEproof}
	The proof follows straight from the definitions.
	\end{IEEEproof}

\subsection{$(P, \pi, L)$-linear isometries}\label{isometriess}

Let $GL(P, \pi, L)_q$ be the group of linear isometries of the space $\Fqn$ endowed with a $(P,\pi, L)$-metric. Our goal in this section is to characterize $GL(P, \pi, L)_q$.

To be more precise, 
\begin{align*}
GL(P, \pi, L)_q &=\{   T: \mathbb{F}_{q}^{n}\rightarrow
\mathbb{F}_{q}^{n}\ : 
\ T \text{ is linear, } \\
 &\hspace{-1cm}d_{(P,\pi,L)}\left(  x,y\right)  =d_{(P,\pi,L)}\left(T  \left( x\right) ,T\left( y\right)  \right) , \forall \ x,y\in \mathbb{F}_{q}^{n} \}\\
&=\{   T: \mathbb{F}_{q}^{n}\rightarrow
\mathbb{F}_{q}^{n}\ : \ T \text{ is linear, } \\ &\hspace{.3cm}\mathrm{wt}_{(P,\pi,L)}\left(  x\right)  =\mathrm{wt}_{(P,\pi,L)}\left(T  \left( x\right)  \right) , \forall \ x\in \mathbb{F}_{q}^{n} \}
\end{align*}

Similarly to what happens in the case of posets, $GL(P, \pi, L)_q$ can be described as the semi-direct product of two subgroups. We start presenting one of them, which is a subgroup of the permutation group $[m]$ that preserves the involved structures: the order structure $P$, the block map $\pi$ and the label function $L$.
\begin{definicao}
	A map $\phi:[m] \rightarrow [m]$ is a $(P, \pi, L)$-automorphism if it is a $P$\emph{-automorphism} with $L(i)= L(\phi(i))$ and $k_i = k_{\phi(i)}$, for every $i\in[m]$. We denote by $Aut(P, \pi, L)$ the set of all $(P, \pi, L)$-automorphisms.
\end{definicao}

We remark that $Aut(P, \pi, L)$ is a group. The following proposition follows straight from the definition of $d_{(P,\pi ,L)}$.

\begin{proposicao}\label{isometrias_automorfismos}
	Let $\phi$ be a $(P, \pi, L)$-automorphism. The linear map $T_{\phi}: \F_q^n \rightarrow \F_q^n$ defined by  $T_{\phi}(e_{ij}) = e_{\phi(i)j}$ is an isometry. Moreover, the map $\varphi: Aut(P, \pi, L)\rightarrow GL(P, \pi, L)$ that associates $\phi \mapsto T_{\phi}$ is an injective homomorphism of groups.
\end{proposicao}

%
%

We denote by $\mathcal{A}:=\{T_{\phi}\in GL(P,\pi,L);\phi \in Aut(P, \pi, L) \}$ the subgroup of isometries \textit{induced by $(P,\pi ,L)$-automorphisms.}
The two next propositions are far from trivial, but the proofs are omitted, for the usual reason.

\begin{proposicao}\label{Normal}
	Let $T: \Fqn \rightarrow \Fqn$ be a linear isomorphism satisfying the following condition: for every $u_i\in \F_q^{k_i}\setminus \{0\}$, there are $u_i'\in \F_q^{k_i}$ and $v_i\in \F_q^{n}$ with $\mathrm{supp}_{\pi}(v_i) \subset \langle i \rangle_P\setminus \{i\}$ such that $T(u_i) = u_i' + v_i$. Then, $T \in GL(P, \pi, L)$.
\end{proposicao}

We denote by $\mathcal{N}$ the set of all the $(P,\pi , L)$-isometries obtained as in Proposition \ref{Normal}. It is possible to prove that $\mathcal{N}$ is a normal subgroup of $GL(P, \pi, L)_q$.
\begin{teorema}\label{theorem2}
	 Every linear isometry $S$ can be written in a unique way as a product $S = F \circ T_{\phi}$, where $F \in \mathcal{N}$ and $\phi \in Aut(P, \pi, L)$. Furthermore, $GL(P, \pi, L)_q$ is the semi-direct product
	$GL(P, \pi, L)_q = \mathcal{N} \rtimes \mathcal{A}.$
\end{teorema}

\subsection{G-Canonical Decomposition of linear codes for hierarchical posets of directed cycles}\label{sec:candec}\label{candec}

Two linear codes $\mathcal{C},\mathcal{C}^{\prime}\subseteq \F_q^n$ are  \emph{$(P,\pi,L)$-equivalent} if there is $T\in GL(P, \pi, L)_q$ such that $T(\mathcal{C})=\mathcal{C}^{\prime}$. 

A decomposition $\mathcal{C}=\mathcal{C}_1\oplus\cdots\oplus\mathcal{C}_{h(P)}$ of a code $\mathcal{C}$ is called  $(P,\pi,L)$\emph{-canonical decomposition} if $\mathrm{supp}_\pi(\mathcal{C}_i)\subseteq\Gamma_i^P$. Working with such decompositions simplifies the computation of all metric invariants of a code. Naturally, not every code admits a $(P,\pi,L)$\emph{-canonical decomposition}, but it may be equivalent to  a code that has such a decomposition.


\begin{definicao}Let $P=([m], \preceq)$ be a poset with $h(P)$ levels. We say that a linear code $\mathcal{C}\subseteq\Fqn$  \emph{admits a} $(P,\pi,L)$\emph{-canonical decomposition} if it is $(P,\pi,L) $-equivalent to a linear code $\tilde{\mathcal{C}}=\mathcal{C}_1\oplus\cdots\oplus\mathcal{C}_{h(P)}$, where   $\mathrm{supp}_\pi(\mathcal{C}_i)\subseteq\Gamma_i^P$.
\end{definicao}

The next theorem is a generalization of the $P$-canonical decomposition for poset metrics, determined in \cite{felix_decomposicao_canonica}.

\begin{teorema}\label{thm:canonical_decomposition}
	The poset $P$ is hierarchical if, and only if, any linear code $\mathcal{D}$ admits a $(P,\pi, L)$-canonical decomposition.
\end{teorema}
\begin{IEEEproof}	This proof can be obtained in a similar way present by Etzion in \cite{etzion}.
\end{IEEEproof}


The existence of a $(P,\pi ,L)$-canonical decomposition is a very useful tool, allowing to simplify the computation of many metric invariants (minimal distance, packing and covering radius) and also to determine conditions which ensure the validity of important results in coding theory, such as the  MacWilliams' Extension Property. Just as an example, we show how it allows to determine a type of  MacWilliams' Identity for linear codes.
%
\begin{definicao}
	A $(P,\pi, L)$-structure satisfies the \textit{unique decomposition property} if, for $1\leq i \leq h(P)$, given $S, S^{\prime}\subseteq \Gamma_i^P$ such that $$ \sum_{a\in S} L(a) = \sum_{b\in S^{\prime}} L(b),$$ there is a bijection $g : S \rightarrow S^{\prime}$ such that $L(a) = L(g(a))$ and $|\pi^{-1}(a)|=|\pi^{-1}(g(a))|$  for all $a\in S$.
\end{definicao}

The $(P,\pi, L)$-\emph{weight enumerator} of a code $\mathcal{C}$ is the polynomial 
$$W^{(P,\pi,L)}_{\mathcal{C}}(X)=\sum_{i=0}^n A^{(P,\pi,L)}_i(\mathcal{C})X^i $$
where $A_i^{(P,\pi,L)}(\mathcal{C})=|\{ c\in \mathcal{C}:\mathrm{wt}_{(P,\pi,L)}(c)=i\}|. $

As we know, given a poset $P([n],\preceq_{{P}})$ its \textit{dual} is the poset $P^\perp([n],\preceq_{P^\perp})$  defined by the opposite relations
$$i \preceq_P j\in E\iff j \preceq_{P^\perp} i.$$
\begin{definicao}(The MacWilliams Identity) A $(P,\pi,L)$-weight \emph{admits} a MacWilliams Identity if for every linear code $\mathcal{C}\subseteq\Fqn$, the $(P,\pi, L)$-{weight enumerator} $W_{\mathcal{C}}^{(P,\pi, L)}(X)$ of $\mathcal{C}$ determines the $(P^\perp,\pi, L)$-weight enumerator $W_{\mathcal{C}^{\perp}}^{(P^\perp,\pi, L)}$ of the dual code $\mathcal{C}^{\perp}$.
\end{definicao}
\begin{teorema} \label{thm:mac_1-level}Consider a $(P,\pi,L)$-weight with $P$ is a hierarchical poset. The $(P,\pi,L)$-weight admits the MacWilliams Identity if, and only if, it satisfies the unique decomposition property.
\end{teorema}

\section*{Acknowledgment}
The authors were supported by grants 2013/25977-7 and 2015/11286-8, S\~{a}o Paulo Research Foundation (FAPESP). 

\bibliographystyle{IEEEtran}
\bibliography{biliog}	

\begin{thebibliography}{1}
\providecommand{\url}[1]{#1}
\csname url@samestyle\endcsname
\providecommand{\newblock}{\relax}
\providecommand{\bibinfo}[2]{#2}
\providecommand{\BIBentrySTDinterwordspacing}{\spaceskip=0pt\relax}
\providecommand{\BIBentryALTinterwordstretchfactor}{4}
\providecommand{\BIBentryALTinterwordspacing}{\spaceskip=\fontdimen2\font plus
\BIBentryALTinterwordstretchfactor\fontdimen3\font minus
  \fontdimen4\font\relax}
\providecommand{\BIBforeignlanguage}[2]{{%
\expandafter\ifx\csname l@#1\endcsname\relax
\typeout{** WARNING: IEEEtran.bst: No hyphenation pattern has been}%
\typeout{** loaded for the language `#1'. Using the pattern for}%
\typeout{** the default language instead.}%
\else
\language=\csname l@#1\endcsname
\fi
#2}}
\providecommand{\BIBdecl}{\relax}
\BIBdecl

\bibitem{deza}
M.~M. Deza and E.~Deza, ``Encyclopedia of distances,'' in \emph{Encyclopedia of
  Distances}.\hskip 1em plus 0.5em minus 0.4em\relax Springer, 2009, pp.
  1--583.

\bibitem{gabidulin2012brief}
E.~Gabidulin, ``A brief survey of metrics in coding theory,'' \emph{Mathematics
  of Distances and Applications}, vol.~66, 2012.

\bibitem{brualdi}
R.~A. Brualdi, J.~S. Graves, and K.~Lawrence, ``Codes with a poset metric,''
  \emph{Discrete Mathematics}, vol. 147, no. 1 - 3, pp. 57 -- 72, 1995.

\bibitem{gabidulin}
E.~Gabidulin, ``Combinatorial metrics in coding theory,'' in \emph{2nd
  International Symposium on Information Theory}.\hskip 1em plus 0.5em minus
  0.4em\relax Akad{\'e}miai Kiad{\'o}, 1973.

\bibitem{muniz}
M.~M.~S. Alves, L.~Panek, M.~Firer \emph{et~al.}, ``Error-block codes and poset
  metrics,'' \emph{Advances in Mathematics of Communications}, 2008.

\bibitem{etzion}
T.~Etzion, M.~Firer, and R.~A. Machado, ``Metrics based on finite directed
  graphs and coding invariants,'' \emph{IEEE Transactions on Information
  Theory}, vol.~PP, no.~99, pp. 1--1, 2017.

\bibitem{rafa}
\BIBentryALTinterwordspacing
R.~G.~L. D'Oliveira and M.~Firer, ``Channel metrization,'' \emph{CoRR}, vol.
  abs/1510.03104, 2015. [Online]. Available:
  \url{http://arxiv.org/abs/1510.03104}
\BIBentrySTDinterwordspacing

\bibitem{felix_decomposicao_canonica}
L.~V. Felix and M.~Firer, ``Canonical- systematic form for codes in
  hierarchical poset metrics,'' \emph{Advances in Mathematics of
  Communications}, vol.~6, p. 315, 2012.

\end{thebibliography}
\end{document}